# Combining Gamification and Intelligent Tutoring Systems in a Serious Game for Engineering Education

Ying Tang
Dept. of Electrical and Computer Engineering
Rowan University
Glassboro, NJ, USA
tang@rowan.edu

Ryan Hare
Dept. of Electrical and Computer Engineering
Rowan University
Glassboro, NJ, USA
harer6@rowan.edu

*Abstract*— **This work-in-progress research-to-practice paper provides ongoing results from the development and testing of a personalized learning system integrated into a serious game. Given limited instructor resources, the use of computerized systems to help tutor students offers a way to provide higher quality education and to improve educational efficacy. Personalized learning systems like the one proposed in this paper offer an accessible solution. Furthermore, by combining such a system with a serious game, students are further engaged in interacting with the system. The proposed learning system combines expert-driven structure and lesson planning with computational intelligence methods and gamification to provide students with a fun and educational experience. As the project is ongoing from past years, numerous design iterations have been made on the system based on feedback from students and classroom observations. Using computational intelligence, the system adaptively provides support to students based on data collected from both their in-game actions and by estimating their emotional state from webcam images. For our evaluation, we focus on student data gathered from in-classroom testing in relevant courses, with both educational efficacy results and student observations.**

**To demonstrate the effect of our proposed system, students in an early electrical engineering course were instructed to interact with the system in place of their standard lab assignments. The system would then measure and help them improve their background knowledge before allowing them to complete the lab assignment. As they played through the game, we observed their interactions with the system to gather insights for future developments, which are presented in this work. Additionally, we demonstrate the system's educational efficacy through early pre-post-test results from students who played the game with and without the personalized learning system integration.**

This work was supported in part by the National Science Foundation under Grant 1913809 and in part by the U.S. Department of Education Gruaduate Assistance in Areas of National Need (GAANN) Grant Number P200A180055.

## I. INTRODUCTION

Despite many recent advancements and paradigm shifts worldwide, classroom education still plays a pivotal role in equipping students with the skills and knowledge required to join the workforce and address real problems. However, larger classrooms can struggle to address student needs and concerns, especially when those needs heavily deviate from the standard one-size-fits-all lesson plan [1, 2]. And as classroom sizes increase, it becomes impossible for instructors to provide support to each individual student based on their needs and background [3]. As such, students often need to resort to excessive studying, self-learning, or external tutoring to gain much-needed knowledge [4].

To take some burden off the instructor while still limiting learning to class time, one possible solution is intelligent tutoring systems (ITSs). ITSs are educational systems that leverage student data to provide personalized educational support and tutoring to students [5]. Unlike human tutors, these systems are not limited by time and resource constraints and can support many students at the same time. By collecting data such as student performance, student actions when interacting with the system, or results on graded assignments, these systems can make informed decisions about what support a student needs to succeed. Furthermore, recent trends in artificial intelligence (AI) and data mining have further bolstered the accuracy and ability of these systems to provide effective support.

However, while these systems can provide appropriate support, they can often struggle with engaging students. This is especially prevalent when students are interacting with systems that require them to read large sections of content, watch extended videos, or otherwise undergo non-active learning [6]. Maintaining this engagement is crucial for effective learning. [7, 8]. The solution to the issue of student engagement, as addressed in this paper, is to combine an ITS with a serious game (SG) [9]. Serious games are virtual or physical games that focus primarily on education, training, or other non-entertainment purposes [10]. Due to the gamification aspect of SGs, students are often significantly more engaged in lessons. In a well-designed game, students may not even realize that they are learning. Additionally, virtual SGs have added benefits in that they can create environments, visualizations, or interactive scenarios that students would not otherwise experience through standard lectures or videos.

This paper presents results from the ongoing development of *Gridlock*, a serious game that focuses on binary logic, simple programming, and digital system design. *Gridlock* is designed to be run in tandem with a lab assignment where students are tasked with designing the logic controller for a traffic light. This task is the goal of the game, and is a common lab assignment for early students in electrical engineering, computer engineering, and computer science. *Gridlock* further supports student learning by combining a serious game environment with an ITS that supports students on various topics relevant to the overarching task. Using reinforcement learning, an AI method, the system automatically adjusts what support it provides based on student responses. This support then manifests as pop-up prompts, hints, adjusted content, and enabling or disabling certain areas in the game based on student performance. We refer to the combined system as the personalized instruction and need-aware gaming (PING) system, and while this paper focuses on Gridlock, the system is designed as a modular, general-purpose approach for any SG.

As the contribution of this work, we provide design insights for both our ITS and our virtual game environment based on both observations of student interactions with the system and interview sessions with students where they provided direct feedback. We also provide educational results from in-classroom testing to verify the educational efficacy of our proposed system. To that end, Section II of this paper gives a more detailed overview of our ITS, *Gridlock*, and the support provided in the game. Section III provides our development insights and educational results, followed by conclusions in Section IV.

## II. OVERVIEW OF PING SYSTEM AND GRIDLOCK

### A. Intelligent Tutoring in the PING System

This section provides a non-technical overview of the PING system and, more specifically, the ITS we integrate for personalized learning. For more technical information, we ask that readers refer to our other recent publication [11, 12]. The PING system is designed as a modular intelligent tutoring system (ITS) that can be integrated with any serious game (SG). To make the ITS modular, the system developer first partitions the game's educational domain into certain subject-specific sections, referred to here as content blocks. Within each content block, the student can then be tested and supported on a specific topic. For example, a game dealing with math might have a block for multiplication and a block for division.

Then, as the student interacts with game components and completes in-game content, the system receives a constant feed of that student's performance, as well as any relevant data. For example, *Gridlock* collects timing information on how long students took to complete sections. In addition, *Gridlock* also records a score on each section, various data on mouse movement and keyboard interaction, as well as estimated emotional response through facial emotion recognition and webcam images. All of this data is collected through quizzes and mini-games, which are covered in the next section.

Then, the data are fed into a set of reinforcement learning agents, each of which is trained to support a student on a specific content block [12]. With reinforcement learning, the system learns optimal behavior by exploring choices, so the system needs some student interaction before providing acceptable student support. But after initial training, the reinforcement learning back-end allows the system to adapt to trends and changes in student behavior without any human intervention.

To learn, reinforcement learning requires feedback about the quality of the decisions made. As such, student performance is used as a "reward" for the agent. If a student performs better after receiving assistance from the system, the system receives a positive reward, encouraging the chosen assistance for future students. Likewise, if a student performs poorly after receiving some assistance, the specific assistance chosen will be less likely to appear to similar students in the future.

### B. Gridlock

*Gridlock* is a game that has been iteratively redesigned and improved over the past several years. Through in-classroom testing, pilot evaluations, and student feedback, numerous changes have been made to improve both the usability of the game and to make the game more enjoyable and educational for students. The game places students in the position to redesign the logic controller for a traffic light, which is a typical lab assignment for courses that teach digital logic design. When the game starts, students are quizzed on their entry knowledge to establish a benchmark for future assistance. Then, students explore a 3D environment and unlock a door by completing tasks specific to the content the game intends to teach.

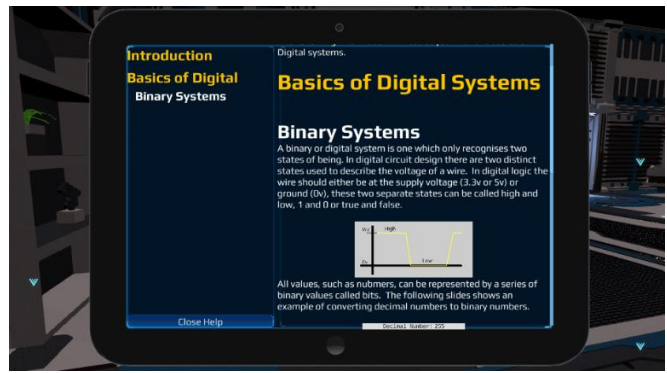

*Figure 1: An excerpt of the help documentation in Gridlock.*

The tasks given to students vary between concepts. For example, one block focuses on the specific design specifications students need for the traffic light problem. In this section, students must complete a miniature design problem which tests their ability to set the order of the traffic light changes. This problem is structured more as a puzzle or a mini-game, and students' number of actions, time taken, and correctness of their submitted solution all give the system insight into the student's knowledge of this section's content. Then, the system can give the student feedback to bolster their content knowledge. As shown in Figure 1, students are given dialogue from the game, videos, images, and other relevant materials based on their performance.

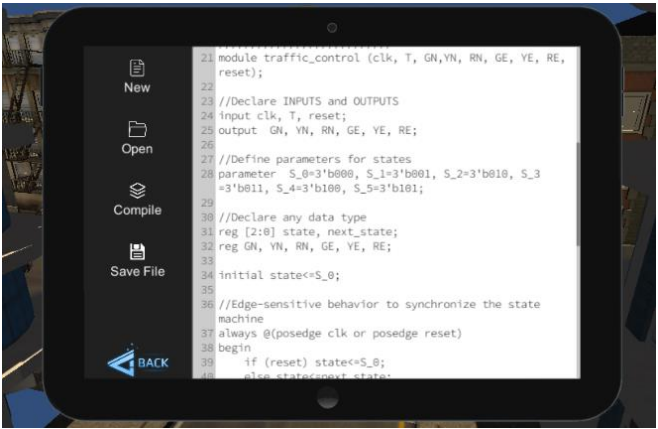

*Figure 2: The in-game coding functionality.*

Once students have mastered the concepts presented by the game, they are tasked with programming their final traffic light logic and submitting it into the game. To help minimize instructor time used by this assignment, the game has full functionality to let students program and debug and simulate their solutions, all within the game, as shown in Figure 2 and Figure 3. With this approach, students have a seamless experience, and the game can observe student performance as they write their solutions.

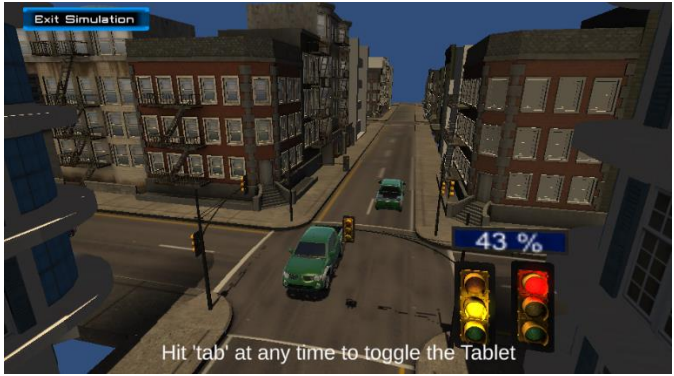

*Figure 3: The traffic light simulator that helps students visualize their solutions.*

### III. Design Insights and Educational Impact

#### A. Design Insights and Student Feedback

When discussing design changes made in the game with regards to student feedback, the iterations focused on three main areas: Interface design, student testing, and tutorials/student assistance.

*1) Interface Design*

For interface design, we had a few main considerations. First, the target audience for the game is highly generalized, and thus includes players who are and are not familiar with virtual games. For players who are familiar with virtual games, certain aspects such as a standard control scheme and familiar interface elements are enough for players to understand the goals, controls, and objective of the game. For example, the updated game features a system of objective markers shown in Figure 4, where students receive a visual indication of their current objective. Even among students who were less familiar with virtual games, a visual indicator marking their objective helps to significantly improve their ability to navigate the virtual environment.

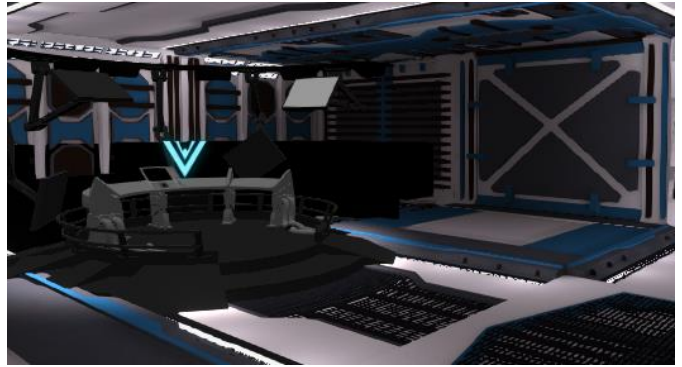

*Figure 4: The waypoints in Gridlock that guide students to their next destination.*

Outside of marking objectives, it was also very important to have both a stated objective and a guide on the control scheme of the game. For *Gridlock*, students can open up a "how to play" screen at any time that indicates the control scheme of the game. Additionally, students can also access a "current objectives" screen that has a written statement of their current objective. Among players experienced with virtual games, these features were largely ignored. However, their inclusion is crucial to help players who are unfamiliar with navigating virtual game environments.

One final aspect that also helped students navigate the environment was "notification" markers on certain in-game utilities. For example, when students start the game, they have a notification on one of the menu options. This incentivizes them to click this option, thus taking them to the assistance menu, as shown in Figure 5. This way, students are already familiarized with this menu before it becomes relevant, allowing them to have a smoother experience in the latter parts of the game.

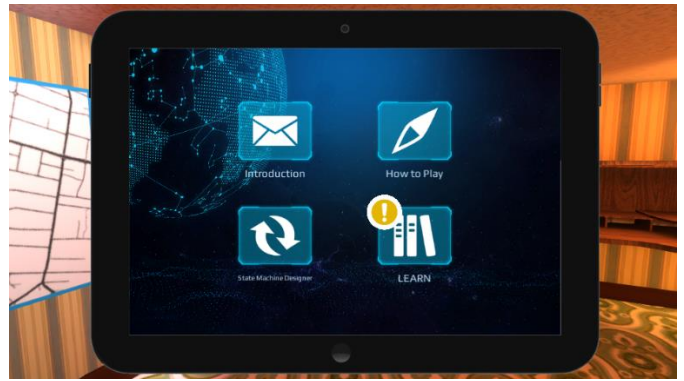

*Figure 5: Another view of the "notification" markers that catch student attention and encourage students to explore certain functions.*

### 2) Student Testing

Another modification that was well-received among students was the replacement of some testing methods with interactive mini-games or puzzles. In each of these basic games, students receive an instructional prompt before participating. These games are designed in such a way that we can gather data on the students' knowledge while they play. For example, as shown in Figure 6, students are tested on their ability to convert binary numbers through a game where they are tasked with matching binary numbers to their decimal counterparts as quickly as possible.

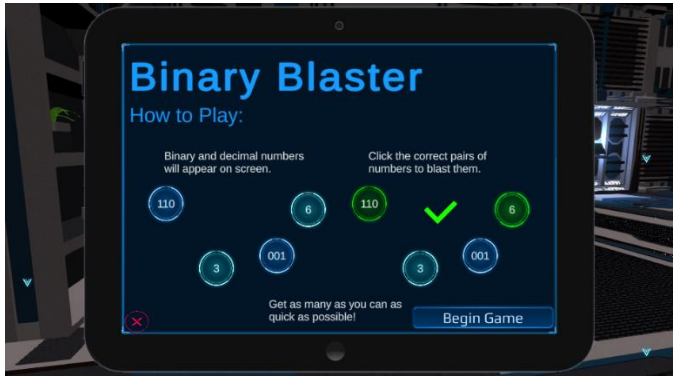

*Figure 6: One of the activities in-game, where students must match binary numbers with their decimal counterparts.*

### 3) Student Assistance

Finally, the in-game tutorials were also modified as per student feedback. In prior iterations, accessing the personalized assistance provided to students was required for students to progress. Furthermore, the assistance was specifically tailored to a student's issues. In the updated version of the game, the assistance is general-purpose, but students are directed to a specific area that addresses their issues. This way, students have the option to manually explore additional help materials. Additionally, students now must manually access the help materials, which prevents the help prompts from breaking the flow of the game. If a student is performing poorly, the game can also put a notification marker on the help menu to encourage the student to request help.

With all these changes, students have much more freedom to either accept the help as given, explore additional content at their own pace, or to ignore the help entirely and attempt to solve problems on their own. This gives students guidance if they need it, and the freedom to explore and solve problems without being limited or interrupted by the system.

### B. Educational Impact

Preliminary results for an evaluation of educational efficacy use a pre-test and a post-test administered before and after students played the game, respectively. These tests asked students to design a state machine for a vending machine; a task which has a similar required skill set as the traffic light problem in the game.

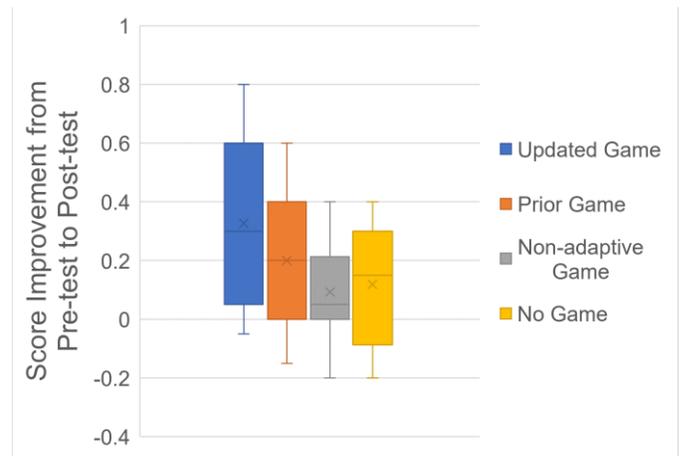

*Figure 7: Comparison of student improvement from pre-test to post-test on three game versions and control group.*

Figure 7 shows a comparison of score improvement among students from our pre-test to the post-test. Students were graded on both assignments on a scale from 0 to 1, and the graph compares four scenarios: 1) The updated game with PING system integration, modified based on student feedback as per the design statements outlined in this paper ($N = 13$); 2) The prior version of the game with PING system integration ($N = 20$); 3) An old version of *Gridlock* that did not use the PING system ($N = 14$); and 4) Students who did not play the game ($N = 8$). As shown, students who played the updated game had greater improvement over students who played the prior version of the game. Additionally, there were a fewer number of students who regressed or showed little improvement compared to past classroom evaluations. This is likely due to the new quality of life features and easier-to-access help documentation making it both easier and less frustrating for students to learn with *Gridlock*.

## IV. CONCLUSION

This paper presents findings from the ongoing development and testing of a personalized intelligent tutoring system integrated into an educational serious game. With our student observation results and design insights, we hope to provide some ideas and direction for other researchers in this area who are interested in creating this type of system. We also show some preliminary results on the educational efficacy of the proposed system. Looking ahead, additional work will be focused on further verifying the efficacy of the system, as well as gathering additional feedback from students and educators to further refine the final product.

## REFERENCES

[1]. D. A. Schwinn, C. S. Cooper, and J. E. Robillard, "putting students at the center: Moving beyond time-variable one-size-fits-all medical education to true individualization," Advances in Medical Education and Practice, vol. Volume 10, pp. 109–112, 2019. doi: 10.2147/amep.s187946

[2]. S. Yang, H. Tian, L. Sun, and X. Yu, "From one-size-fits-all teaching to adaptive learning: The crisis and solution of education in the era of ai," Journal of Physics: Conference Series, vol. 1237, no. 4, p. 042039, 2019. doi: 10.1088/1742-6596/1237/4/042039

[3]. S. Olson, Grand Challenges for Engineering Imperatives, Prospects, and Priorities: Summary of A Forum. Washington, D.C: National Academies Press, 2016.

[4]. T.-C. Lin, "An investigation of the relationship between in-class and out-of-class efforts on student learning: Empirical evidence and strategy suggestion," Journal of the Scholarship of Teaching and Learning, vol. 16, no. 4, pp. 14–32, 2016. doi: 10.14434/josotl.v16i4.20028

[5]. E. Mousavinasab et al., "Intelligent tutoring systems: A systematic review of characteristics, applications, and evaluation methods," Interactive Learning Environments, vol. 29, no. 1, pp. 142–163, 2018. doi: 10.1080/10494820.2018.1558257

[6]. F. Martin and D. U. Bolliger, "Engagement matters: Student perceptions on the importance of engagement  strategies in the Online Learning Environment," Online Learning, vol. 22, no. 1, 2018. doi: 10.24059/olj.v22i1.1092

[7]. G. Hookham and K. Nesbitt, "A systematic review of the definition and measurement of engagement in serious games," Proceedings of the Australasian Computer Science Week Multiconference, 2019. doi: 10.1145/3290688.3290747

[8]. M.-T. Wang and T. L. Hofkens, "Beyond classroom academics: A school-wide and multi-contextual perspective on student engagement in school," Adolescent Research Review, vol. 5, no. 4, pp. 419–433, 2019. doi: 10.1007/s40894-019-00115-z

[9]. M. Beyyoudh, M. K. Idrissi, and S. Bennani, "A new approach of integrating serious games in intelligent tutoring systems," Learning and Analytics in Intelligent Systems, pp. 85–91, 2019. doi: 10.1007/978-3-030-36778-7_10

[10]. F. Laamarti, M. Eid, and A. El Saddik, "An overview of serious games," International Journal of Computer Games Technology, vol. 2014, pp. 1–15, 2014. doi: 10.1155/2014/358152

[11]. J. Liang, Y. Tang, R. Hare, B. Wu and F. -Y. Wang, "A Learning-Embedded Attributed Petri Net to Optimize Student Learning in a Serious Game," in IEEE Transactions on Computational Social Systems, 2022, doi: 10.1109/TCSS.2021.3132355.

[12]. R. Hare and Y. Tang, "Hierarchical Deep Reinforcement Learning With Experience Sharing for Metaverse in Education," in *IEEE Transactions on Systems, Man, and Cybernetics: Systems*, vol. 53, no. 4, pp. 2047-2055, April 2023, doi: 10.1109/TSMC.2022.3227919.